\documentclass[fleqn,usenatbib]{mnras}

\usepackage{newtxtext,newtxmath}
\usepackage[T1]{fontenc}
\usepackage{ae,aecompl}

\usepackage[pdftex]{graphicx}	% Including figure files
\usepackage{amsmath}	% Advanced maths commands
\usepackage{amssymb}	% Extra maths symbols
\title[Survival of closely-packed systems]{Survival of non-coplanar, closely-packed planetary systems after a close encounter}

\author[D. R. Rice et al.]{
David R. Rice,$^{1,2}$\thanks{E-mail: david.rice@unlv.edu}
Frederic A. Rasio,$^{2}$
and Jason H. Steffen$^{1}$
\\
$^{1}$Department of Physics \& Astronomy, University of Nevada, Las Vegas, 4505 S. Maryland Pkwy., Las Vegas, NV 89154, USA\\
$^{2}$CIERA and Department of Physics \& Astronomy, Northwestern University, 2145 Sheridan Road, Evanston, IL 60208, USA
}

\date{Accepted XXX. Received YYY; in original form ZZZ}

\pubyear{2018}

\begin{document}
\label{firstpage}
\pagerange{\pageref{firstpage}--\pageref{lastpage}}
\maketitle

% Abstract of the paper
\begin{abstract}
Planetary systems with more than two bodies will experience orbital crossings at a time related to the initial orbital separations of the planets.  After a crossing, the system enters a period of chaotic evolution ending in the reshaping of the system's architecture via planetary collisions or ejections.  We carry out N-body integrations on a large number of systems with equally-spaced planets (in units of the Hill radius) to determine the distribution of instability times for a given planet separation.  We investigate both the time to the initiation of instability through a close encounter and the time to a planet-planet collision.  We find that a significant portion of systems with non-zero mutual inclinations survive after a close encounter and do not promptly experience a planet-planet collision.  Systems with significant inclinations can continue to evolve for over 1,000 times longer than the encounter time.  The fraction of long lived systems is dependent on the absolute system scale and the initial inclination of the planets.  These results have implications to the assumed stability of observed planetary systems.
\end{abstract}

% Select between one and six entries from the list of approved keywords.
% Don't make up new ones.

%%%%%%%%%%%%%%%%%%%%%%%%%%%%%%%%%%%%%%%%%%%%%%%%%%

%%%%%%%%%%%%%%%%% BODY OF PAPER %%%%%%%%%%%%%%%%%%

\section{Introduction}\label{sec:intro}

Through various observational techniques, 628 multi-planet exoplanetary systems have been confirmed\footnote{exoplanets.eu, as of May 1, 2018}.  Many of the planets in these systems orbit in close proximity to each other.  Examples of compact systems are Kepler-11 with six planets within 0.5 AU of a G-type star \citep{lissauer} and TRAPPIST-1 with seven planets within 0.06 AU of an M-dwarf star \citep{gillon}.  Most of the high-multiplicity systems are ``dynamically packed'', so that an additional planet would be unstable \citep{fang}.  Both \citet{pu} and \citet{volk} show that dynamical instabilities can clear out planetary embryos that are initially even more packed to form the observed systems.  In the post-gas disk phase, eccentricities of embryos will grow through gravitational perturbations until their orbits cross.  When the bodies encounter one another the system enters a time of chaotic evolution.

A planetary system with only two bodies can be strictly stable when the difference between the semi-major axes exceeds 2 $\sqrt[]{3}$ times their mutual Hill radius \citep{gladman}.  The mutual Hill radius is defined as
\begin{equation}
R_H = [(m_1 + m_2)/3M]^{1/3}[(a_1 + a_2)/2], \label{eq:hill}
\end{equation}
where $m_1$ and $m_2$ are the planetary masses, $a_1$ and $a_2$ are their semi-major axes, and $M$ is the mass of the central body.  Consequently, planet separation can be defined in terms of a spacing parameter, $\Delta$, as
\begin{equation}
a_2-a_1=\Delta R_{H}.\label{eq:sep}
\end{equation}

In systems with more than two planetary bodies the energy and angular momentum of a given planet pair are not conserved because of perturbations from the additional planets.  This results in the orbits of the planets eventually crossing one another, even in systems with initially large separations.  \citet{chambers} is one of the first to study these complex interactions as they pertain to multi-body systems.  Through orbital calculations of equal-mass protoplanets on initially circular and coplanar orbits, they find an exponential relationship between the orbital spacing and the time from initial conditions to the first close encounter (defined as a separation of less than one mutual Hill Radius).  We refer to this time as the ``encounter time''.  The empirical relationship is given by
\begin{equation}
\log(t)= b(\Delta) + c.\label{eq:chamber}
\end{equation}
The values of the constants depend on planet mass, multiplicity, eccentricity and inclination \citep{chambers, yoshinaga}.

Numerous studies explore this relationship but limit their analysis by equating the ``instability time'' of the system with the encounter time \citep{veras, smith, pu}.  Other studies further limit the parameter space by analyzing only coplanar systems \citep{zhou, faber, shikita, matsumoto, morrison, obertas}.  \citet{ford1} and \citet{ford} investigate the effect of these two constraints in two planet systems.  For the case of non-coplanar, multi-body systems, the timescales of system-shaping events are analyzed in the specific case of Kepler-11 by \citet{hwang} and in close-in systems of $a<0.15$ AU by \citet{petrovich}.  Additionally, \citet{chatterjee} and \citet{dawson2} detail the final orbital properties of planets and embryos after a planet-planet scatterings and collisions.

Here, we investigate non-coplanar, equally-spaced, multi-body systems by simulating a large number of idealized planetary systems with varying semi-major axes and mutual inclinations.  We extend the results of \citet{petrovich} out to 100 AU (beyond the regime where the planet-planet encounter energy is comparable to the planet-star binding energy).  We investigate not only the timescale of  close encounters but also the timescale of a planet-planet collision.  By using non-coplanar systems, we find significant differences from previous coplanar studies on the timescales of instabilities depending upon how the instability time is defined.

Our paper is laid out as follows.  In Section \ref{sec:methods} we outline the setup of our near-coplanar systems and show our agreement with Chambers' encounter time.  We then examine at a given orbital separation the distribution of encounter times in Section \ref{sec:encounter}.  In Section \ref{sec:collision} we increase the integration time of our simulations to analyze the time from initial conditions to the first collision between a pair of planets (referred to from here on as ``collision time'').  In Section \ref{sec:evolution} we vary the initial inclinations and eccentricities given to the planets and look at the evolution of those orbital elements.  Determining which planets are involved in the instability events is investigated in Section \ref{sec:branches}.  Finally, we present our conclusions and discuss the implications of this work in Section \ref{sec:conclusion}.

\section{Simulations}\label{sec:methods}

We use N-body integrations to evolve planetary systems in order to study the timescales over which instability is manifested.  Our simulations use the Bulirsch-Stoer (B-S) Method in the software package \textit{Mercury6.2}\footnote{\textit{Mercury6} can be found at http://www.arm.ac.uk/~jec/home.html} \citep{chambers2}.  The accuracy parameter is kept at 10$^{-12}$.  The initial time-step is always set at a time less than 1/20\textsuperscript{th} of the innermost planet's period.  The central body has a mass of 1.0 $M_{\odot}$ and a radius of 0.005 AU throughout the study.

We use suites of 1,000 simulations, each containing four Neptune-like planets.  Each planet has a mass of 10$^{-5}$ $M_{\odot}$ and a density of 2.00 g/cm$^{3}$.   We start our systems with the planets spaced by a constant spacing parameter.  The innermost planet is placed at 1.0 AU while subsequent planets' semi-major axes are determined by the orbital separation imposed on the system.  Specifically, from Equations (\ref{eq:hill}) and (\ref{eq:sep}), each subsequent planet's semi-major axis is
\begin{equation}
a_{2} = a_{1}(2 + \Delta K)/(2 - \Delta K),\label{eq:semi}
\end{equation}
where $K$ for planets of equal mass, $m$, is \((2m/3M)^{1/3}\).  We do not consider atmospheric interactions of Neptune-like planets \citep{hwang1} or the observed orbital parameters of similar exoplanets \citep{mazeh}.  The initial choices are made to keep instability times short and the effects of mass and radius apparent.

We choose inclinations and eccentricities from Rayleigh distributions, which \citet{fang} showed to be the approximate distribution of observed systems.  For our first study we use systems that are near-circular and  near-coplanar to be comparable with \citet{chambers}.  We use a Rayleigh scale parameter of 10$^{-5}\cdot\sqrt[]{2/\pi}$ giving us random values between 10$^{-6}$ and 10$^{-4}$.  For each planet the argument of pericenter, longitude of the ascending node, and the mean anomaly are chosen randomly from $0-360$ degrees.  The planets are given no spin angular momentum.  Initial conditions of the planets and their orbits are summarized in Table \ref{tab:IC}. 

\begin{table*}
	\centering
	\caption{Initial conditions for each of the four planets in our first simulations.\label{tab:IC}}
	\begin{tabular}{lccr} % four columns, alignment for each
		\hline
		 & Value/Values & Details\\
		\hline
		Mass ($M_{\odot}$)& 0.00001 & fixed \\
		Density (g/cm$^{3}$) & 2.00 & fixed \\
		Semi-Major Axis (AU) & 1.00 - 1.33 \quad & Eq.(\ref{eq:semi}) Values show range for $\Delta$=5  \\ 
		Eccentricity & $\approx$10$^{-6}$-10$^{-4}$ & Rayleigh random around 10$^{-5}$ \\ 
		Inclination ($^{\circ}$) & $\approx$10$^{-6}$-10$^{-4}$ &Rayleigh random around 10$^{-5}$  \\ 
		Arg. of Pericenter ($^{\circ}$) &0 - 360 & Random \\ 
		Long. of Ascend. Node ($^{\circ}$) &0 - 360 & Random \\ 
		Mean Anomaly ($^{\circ}$) &0 - 360 & Random \\ 
		\hline
	\end{tabular}
\end{table*}

\begin{figure}
	\includegraphics[width=\columnwidth]{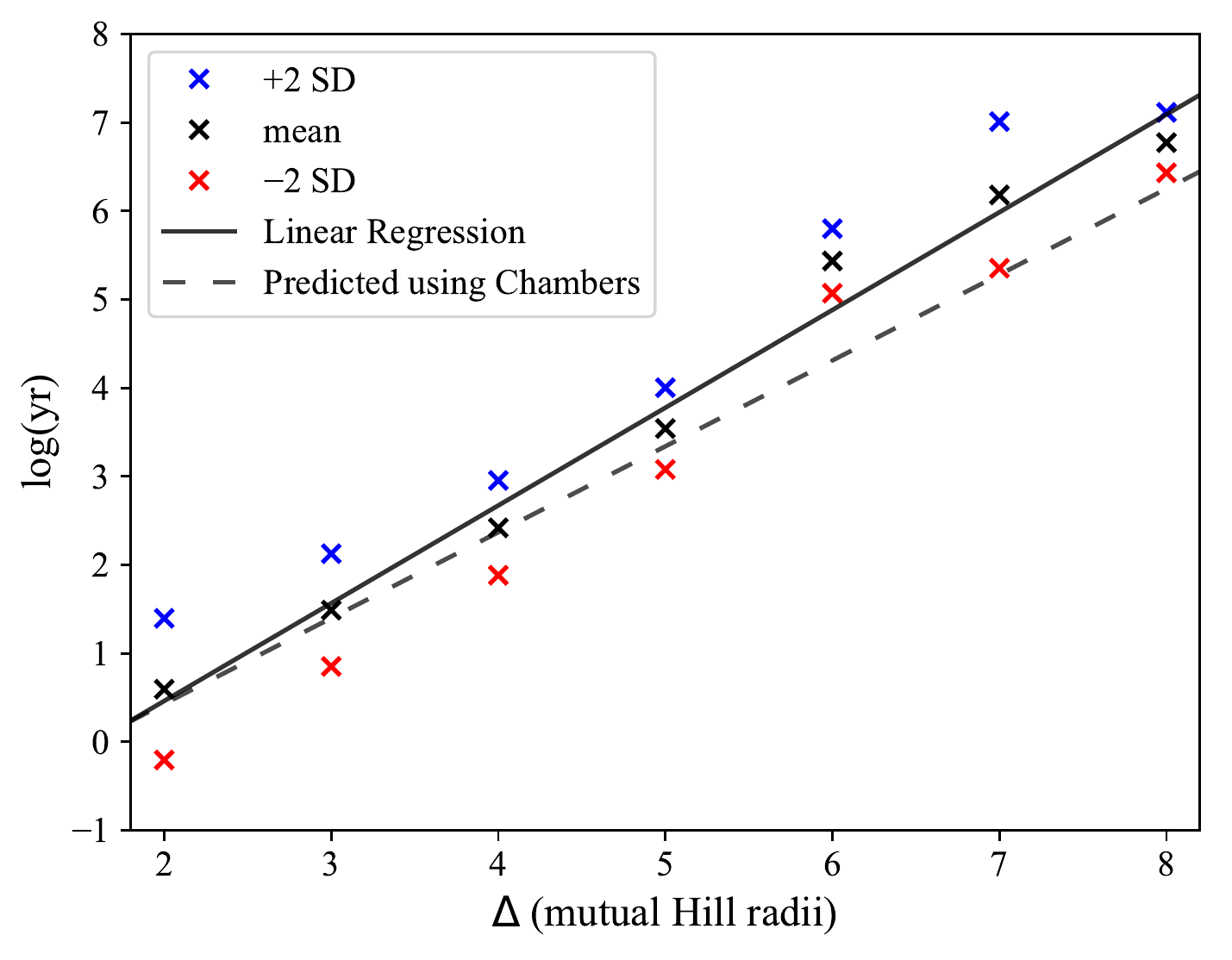}
	\caption{1,000 systems of equally spaced planets are simulated at each integer orbital separation (\(\Delta\)) from 2 to 8 mutual Hill radii.  Each system contains four 10$^{-5} M_{\odot}$ planets on near-circular and near-coplanar orbits.  We show in log-time the mean and $\pm 2\sigma$ time from initial conditions to the first close-encounter of less than one Hill Radius for each integer spacing.  Our exponential relationship between orbital separation and close encounter time is compared to a predicted relationship from averaging the linear regressions reported in \citet{chambers} with synodic period correction.  \label{fig:deltas}}
\end{figure}

As an initial test, we investigate the relationship between orbital spacing and encounter time.  A close encounter in our simulations is a planet conjunction of less than one Hill Radius.  We run suites of simulations with the above initial conditions at integer orbital spacing between $2 \leq \Delta \leq 8$ with the innermost planet at 1.0 AU.  Fig. \ref{fig:deltas} shows the resulting exponential relationship.  A least-squares fit to the data results in a slope of 1.1047 $\pm$ 0.0024 and an intercept of -1.7479 $\pm$ 0.0124 with a correlation coefficient of 0.983.  

We compare our work to \citet{chambers} by predicting a relationship for our systems.  Since \citet{chambers} does not consider four-planet systems, We average the reported least-squares fit for systems of three and five planets each with mass of $10^{-7} M_{\odot}$.  As discussed in \citet{chambers}, the mass of the planets primarily influences the intercept while the slope only has a small dependency on mass.  We correct the interpolated relationship from $10^{-7} M_{\odot}$ planets to $10^{-5} M_{\odot}$ planets by applying the synodic period correction to the intercept found in \citet{chambers}.  The predicted relationship has a slope of 0.971 $\pm$ 0.058 and an intercept of -1.513 $\pm$ 0.328.  Our data is within the error of the predicted intercept.  The slope of our data is higher than predicted, but by averaging the three and five multiplicity systems we assumed that the slope varied linearly with planet multiplicity, which is not expected.  Some of the more detailed structure of the relationship is due to nearby mean-motion resonances (MMR) which is explored in detail in \citet{obertas}.

\section{Timescale to First Encounter}\label{sec:encounter}

\begin{figure}
\includegraphics[width=\columnwidth]{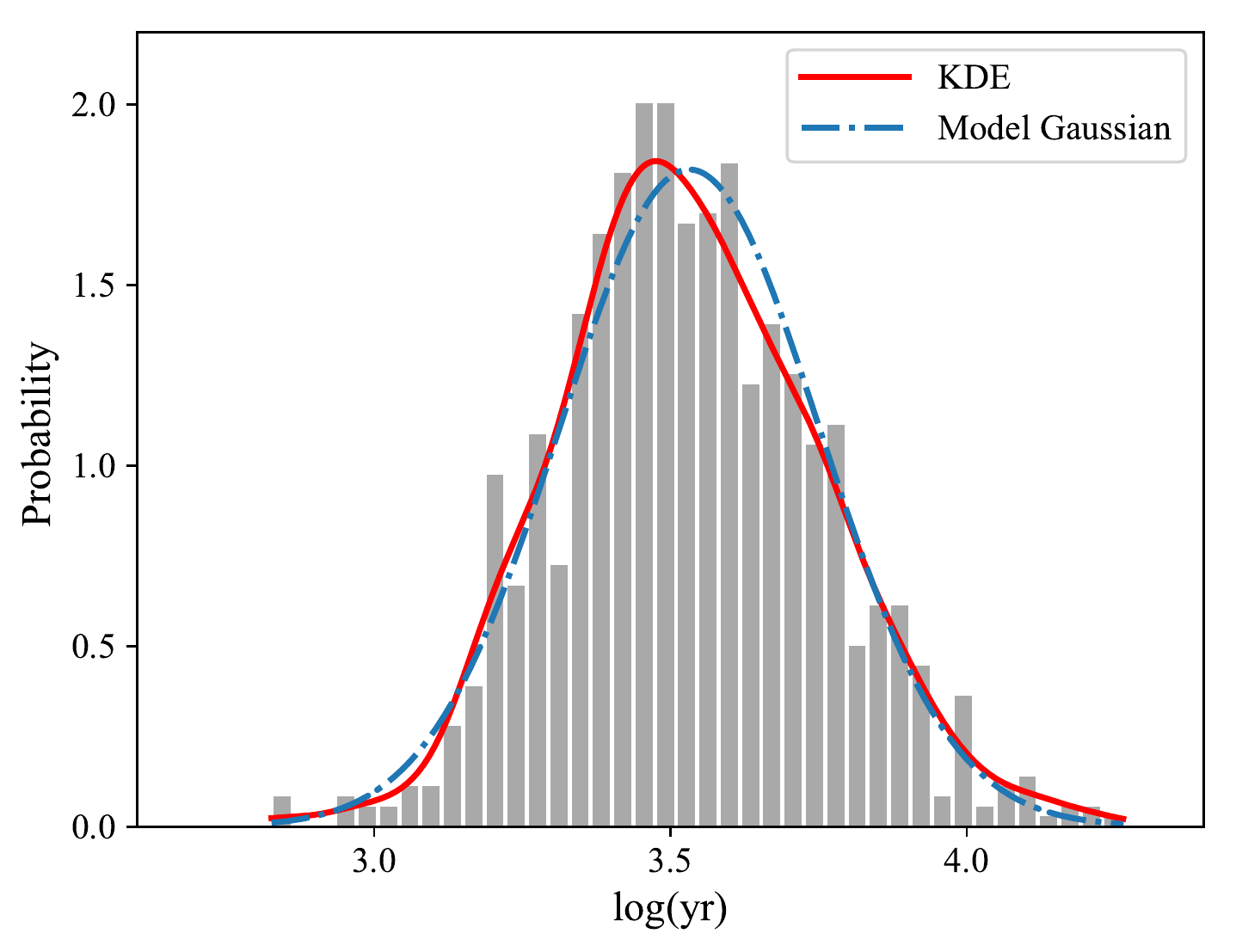}
\caption{The probability density function of encounter times for equally spaced systems of $\Delta=5$ shown as a normalized histogram and as smoothed by a Gaussian kernel density estimator (KDE).  The KDE has a bandwidth determined by Scott's rule namely, \(n^{-1/(d+4)}\), where $n$ is the number of data points and $d$ is the number of dimensions \citep{scott}.  The KDE is compared to a model normal distribution using the sample mean and standard deviation in log-space. \label{fig:gauss}}
\end{figure}

With our suites of simulations at each integer $\Delta$, we look at the distribution of encounter times in each suite.  This encounter timescale distribution was first analyzed by \citet{chatterjee}.  We initially analyzed several suites, but choose to consider the $\Delta=5$ suite in detail throughout the rest this study.  Fig. \ref{fig:gauss} shows the probability density function for our near-circular and -coplanar systems. The distribution of encounter times is shown to be log-normal by a normality test with a p-value of \(<0.05\) \citep{normality}.  The distribution has a mean of 3.53 log-yrs and standard deviation of 0.219 log-yrs.  

\begin{figure}
\includegraphics[width=\columnwidth]{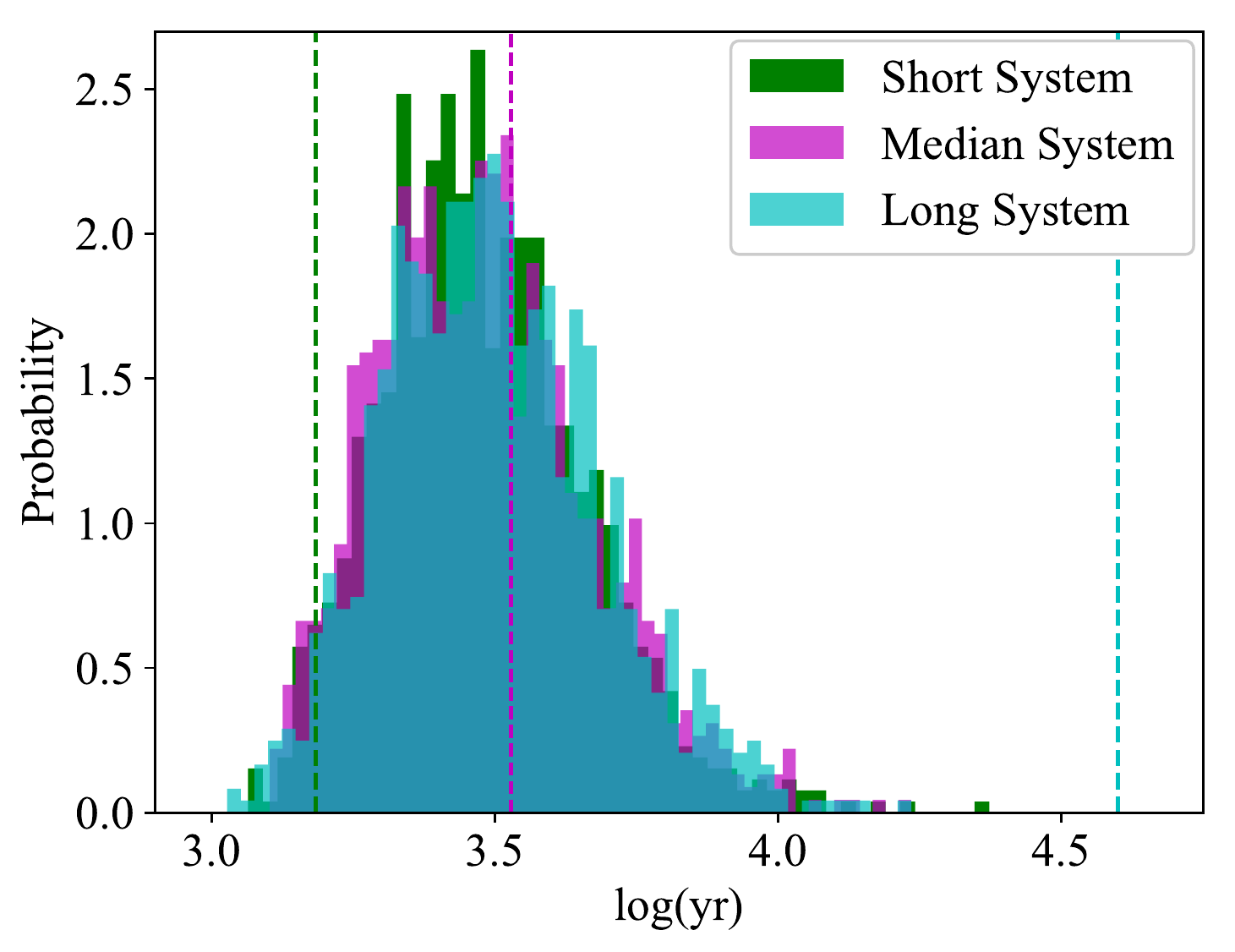}
\caption{Normalized histograms of encounter times for the three suites of replicas created from the chosen systems.  Dashed lines mark the encounter time of the original ``short'', ``median'', and ``long'' system chosen.  All 1,000 systems in each suite have identical initial conditions to the chosen system except for a small change in the argument of periapsis of one random planet.  The change is made by choosing a random number from a normal distribution around the original argument of periapsis with standard deviation of $10^{-4}$ degrees. \label{fig:variation}}
\end{figure}

We test the ubiquity of the encounter timescale distribution by making small variations to a system.  From the original 1,000 systems we choose three systems across the distribution: one with a short encounter time, one near the median time, and one with a long encounter time.  We create 1,000 replicas of each of those systems.  In each replica, we change the argument of periapsis of one randomly selected planet by adding a normally distributed random variable with standard deviation of $10^{-4}$ degrees.  After making this change, the original distribution of encounter times was recovered for all three suites (Fig. \ref{fig:variation}).  The distribution of encounter times for systems with small differences expands to become virtually identical to the distribution for the initial systems\footnote{We note that an ensemble of these random system variates will yield a small fraction that do not reproduce the initial distribution---especially very near its extremes.  This situation is likely due to the effects of resonances or encounters very near the start of the simulation.}.

\begin{figure}
\includegraphics[width=\columnwidth]{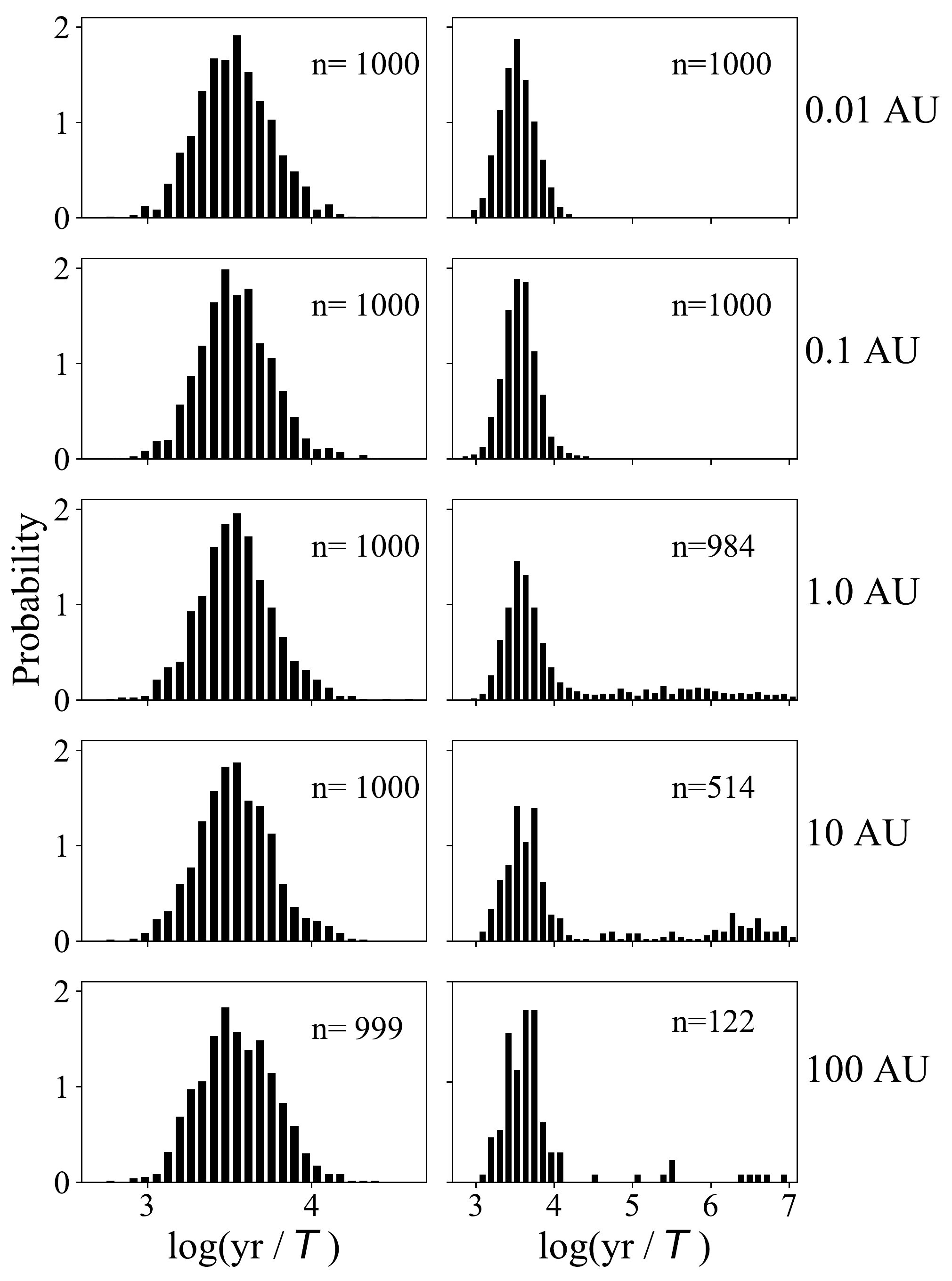}
\caption{\textit{Left}, normalized histograms of encounter times for 1,000 systems with $\Delta = 5$ and four near-circular and near-coplanar planets.  \textit{Right}, normalized histograms of collision times.  Time is transformed into a dimensionless unit by dividing by the period of the innermost planet at it's initial position.  Systems from top to bottom are scaled by changing the innermost planet's semi-major axis in multiples of 10 from 0.01 AU to 100 AU (noted on the right).  The number of systems having an event within the integration time is labeled `\textit{n}'.  One system had an error in the 100 AU suite and was left out.\label{fig:collisioncompare}}
\end{figure}

Since the encounter time depends upon the planet separations as measured in mutual Hill radii, and the Hill radii are proportional to the semi-major axes, we expect the distribution of encounter times to be independent of the scale of the system---the initial semi-major axes of the planetary orbits---at least until some other physical scale becomes relevant to the dynamics.  To test this assumption, we run suites with equal-spacing of $\Delta=5$, but scaling the system by placing the innermost planet at 0.01, 0.1, 1.0, 10, and 100 AU.  The results are shown in the left set of panels in Fig. \ref{fig:collisioncompare}.  The orbital period of the planets grows with system scale, so to compare the distribution shapes we measure time in orbits of the innermost planet at its initial position.  In this dimensionless unit, the distributions are visually similar.  The percent difference in mean encounter time between the 0.01 AU systems and 100 AU is only 3.68\%.

%However, at large system scales, the escape velocity from the system becomes less than the escape velocity from the planets with $R_{p}\approx 13340$ km.  This occurs at a distance of 
%\begin{equation}
%a=\frac{M}{m_{p}}R_{p}\approx 8.917\text{ AU},
%\end{equation}
%for planets with $m_{p}=10^{-5} M_{\odot}$ and $R_{p}\approx 13340$ km. Since the maximum amount of energy a planet can gain from a close encounter is dependent on the escape velocity from the planet, 

\section{Encounter to Collision}\label{sec:collision}

After planets undergo a close encounter, the system enters an era of chaotic evolution often marked by large changes in the orbits of the planets.  The ultimate manifestation of the system's instability is a planet-planet collision, planet-sun collision, or planet ejection.  By continuing the integration of our $\Delta=5$ simulations, we compare the difference in the distributions of encounter times and collision times.  We use a simple definition for planetary collisions by recording when the planet's radii cross.  The six suites used in the previous section, Section \ref{sec:encounter}, are continued to the first planetary collision.  The maximum integration time is set to $10^{7} $orbits of the innermost planet at its initial position---three orders of magnitude longer than the latest first encounter and long enough to identify trends (though not all systems have collisions in that time).  No ejections or collisions with the central body are observed over all suites.  

\begin{figure}
\includegraphics[width=\columnwidth]{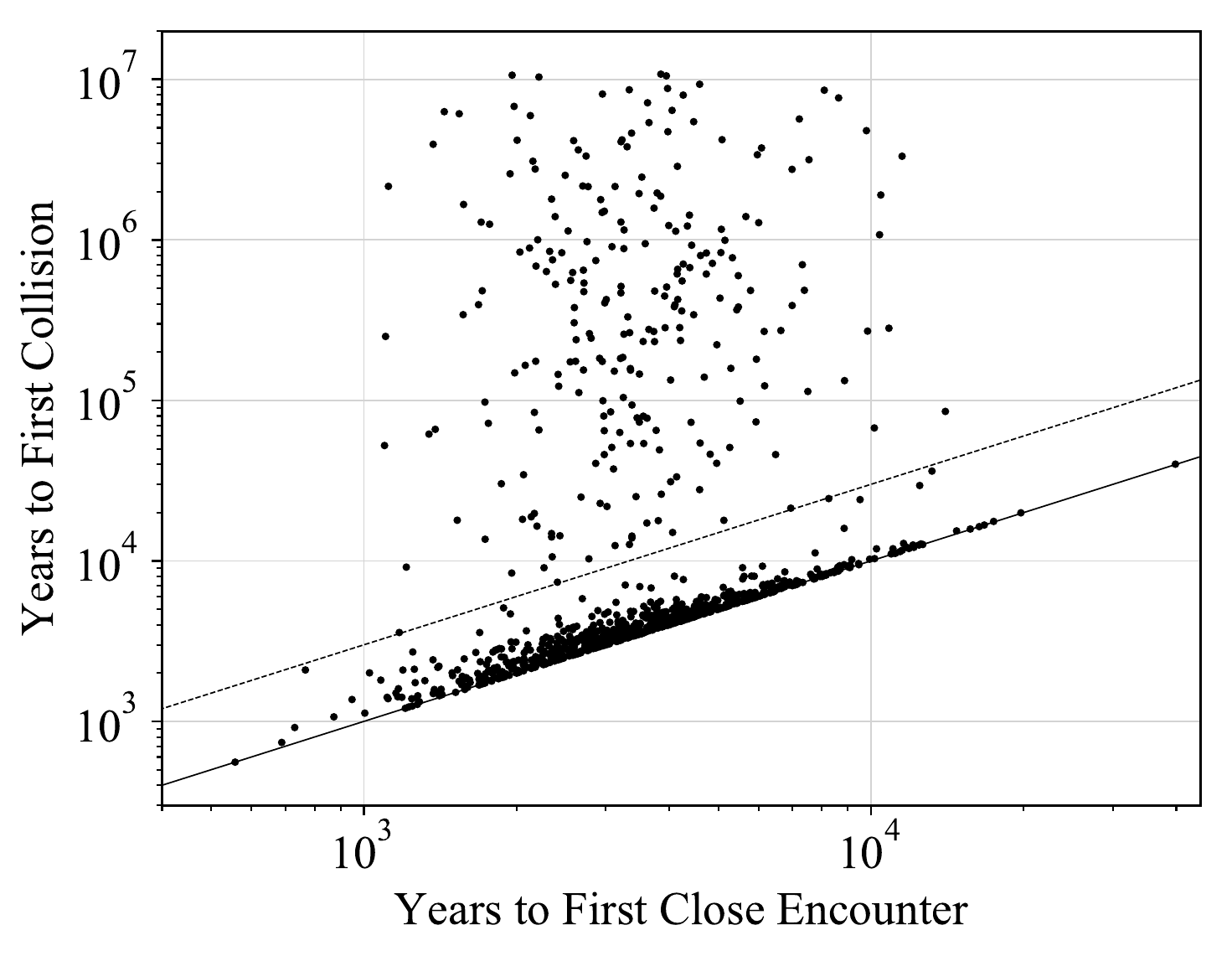}
\caption{A comparison of each system's time from initial conditions to a planet-planet collision versus the encounter time for the suite of 1,000 system with $\Delta=5$ and the innermost planet at 1 AU.  The lower solid lines marks when the collision and encounter are simultaneous. The upper dashed line marks our chosen threshold between prompt and long lived systems, $t_{col}/t_{enc}= 3$.  A majority of systems have a prompt collision soon after the close encounter.  16 systems that were without a collision within the integration time have been removed. \label{fig:scatter}}
\end{figure}

We find that a portion of systems evolve without a collision for a long period of time after their first encounter.  In the 1.0 AU, $\Delta=5$ suite seen in Fig. \ref{fig:scatter} about 72\% of systems follow the encounter time distribution.  We describe these as having a ``prompt'' collision, which for our purposes we define as $t_{col}/t_{enc}\le 3$ where  $t_{col}/t_{enc}$ is the ratio of the collision time to the encounter time.  The remaining $\simeq28\%$ of systems are ``long lived'' and have collision times that are broadly distributed across the duration of the simulations.  Sixteen of the one thousand systems do not have a collision within the integration time, and only five of those systems collide if the integrations are extended to $10^{7.5}$ orbits.  The distribution of collision times is recovered when making perturbations to select systems as done in Fig. \ref{fig:variation} for the distribution of encounter times.

Although the absolute scale of the system---as characterized by the semi-major axis of the innermost planet---was shown to have no effect on the distribution of encounter times (Section \ref{sec:encounter}), the average collision time increases with the scale of the system (Fig. \ref{fig:collisioncompare} \textit{Right}).  With the innermost planet at 0.01 AU the distribution of collision times is virtually identical to the distribution of the encounter times.  At larger scales, however, the majority of systems become long lived systems with $t_{col}/t_{enc}\ge 3$.  In the suite with the innermost planet at 100 AU only 12\% of systems have a planet-planet collision within our integration time of $10^{7}$ orbits. 

For a system of low-$\Delta$ to ``survive'' for a relatively long period of time following the initial encounter, it must be experiencing additional encounters that do not cross to within the radii of the planets.  When the innermost planet is at 0.01 AU, the planet's radius is close to 60\% of it's Hill Radius and collisions occur promptly.  While, at large scales, the radius of the Hill Sphere is much larger. Thus the probability is lower that a close encounter of less than one Hill Radius is also within the radius of the planet.  With a lower collision probability the system has longer to evolve before a collision. We see in the following section how this evolution leads to long lived systems.

\section{Evolution of Eccentricity and Inclination}
\label{sec:evolution}

\begin{figure}
\includegraphics[width=\columnwidth]{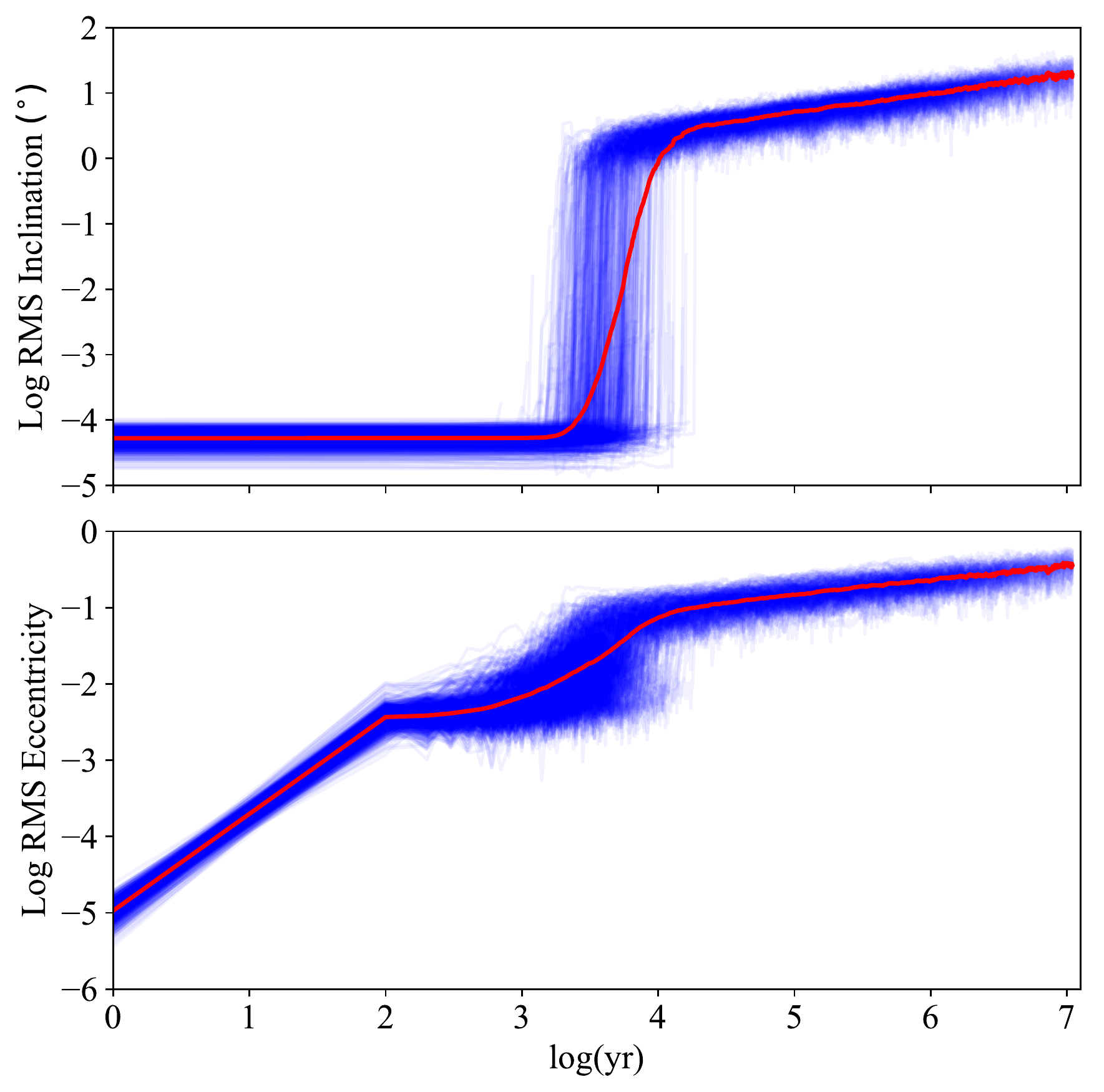}
\caption{The evolution of inclination, \textit{top}, and eccentricity, \textit{bottom}, for each of the 1,000 near-coplanar systems is shown on a log-log scale.  Each system is represented by a thin, blue line that traces the root mean square value of the four planets in the system.  Inclination is measured in degrees from the initial ecliptic plane.  The mean inclination/eccentricity of surviving systems is shown as a thick, red line.  Orbital elements are recorded every 100 years. \label{fig:loginc}}
\end{figure}

We now investigate the dynamical evolution of the long-lived systems that survive following a close encounter.  We analyzed the evolution of inclination and eccentricity in our suite of systems with the innermost planet at 1 AU and $\Delta=5$.  Initially, the root mean square (RMS) eccentricity of the four planets in each simulation has a typical values of $10^{-5}$, which grows rapidly (less than our shortest recorded time of one hundred orbits) to a quasi-equilibrium value of $\sim 10^{-2}$ (see Fig. \ref{fig:loginc}).  Over time, the eccentricity distribution spreads such that its tail reaches large enough eccentricities for close encounters between planets to occur.  As the encounters begin, the RMS eccentricities transition to a new evolutionary path where they continue to grow with a power-law form of approximately $e \propto t^{1/6}$.  Such behavior cannot persist indefinitely given the maximum eccentricity of bound orbits.  However, it does persist over at least three orders of magnitude in time---during which the mean eccentricity grows from $\sim 0.1$ to $\sim 0.3$.

%\subsection{Inclination}\label{sec:inclination}

\begin{figure}
\includegraphics[width=\columnwidth]{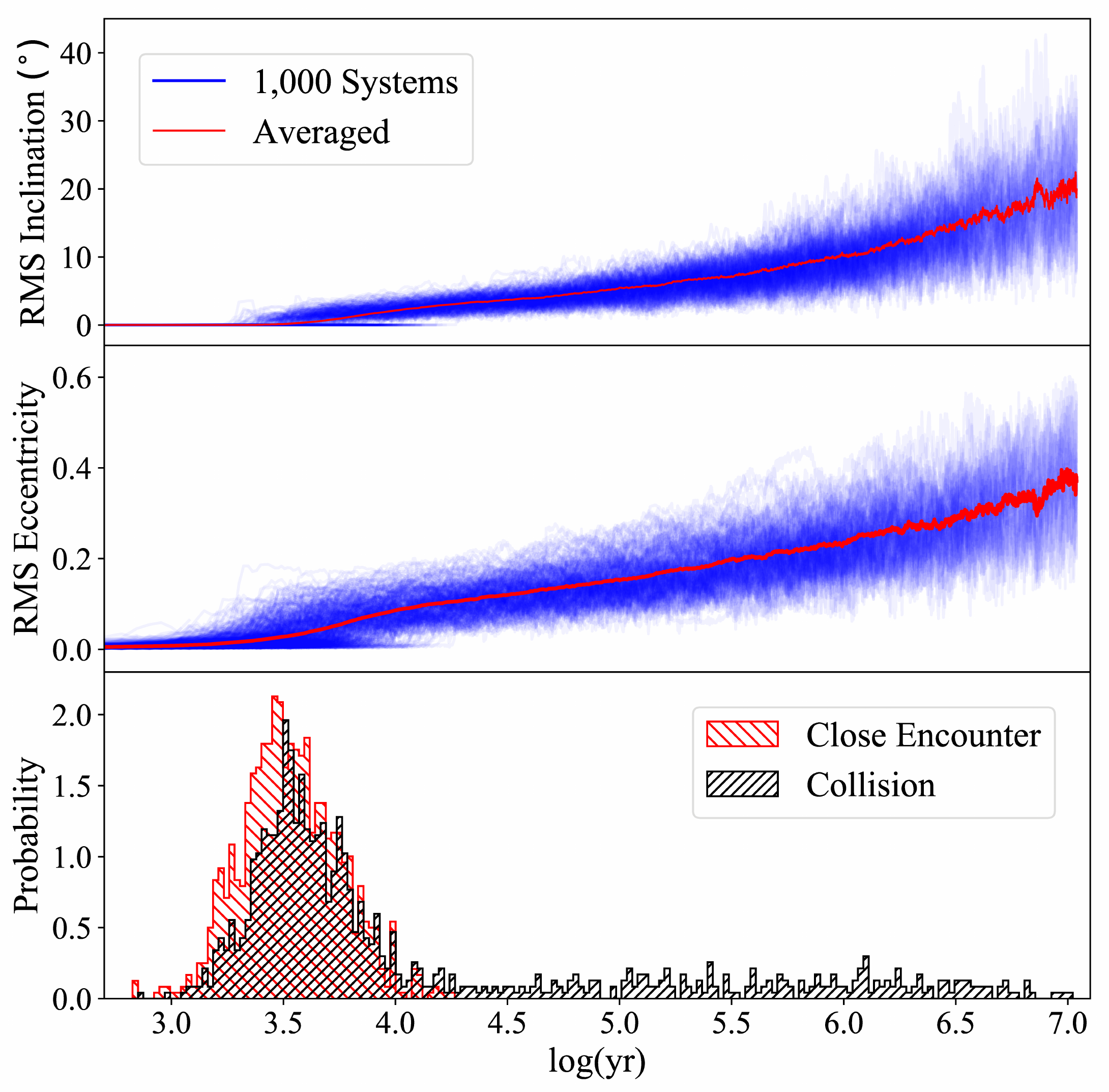}
\caption{The evolution of RMS inclination, \textit{top}, and RMS eccentricity, \textit{middle}, for the four planets in each 1,000 near-coplanar systems is shown on a linear-log scale.  Inclination is measured in degrees from the initial ecliptic plane.  \textit{Bottom}, normalized histograms of both encounter times and collision times for comparison of when events occurred.  \label{fig:lininc}}
\end{figure}

In conjunction with the eccentricities, the RMS inclination in a system also increases.  One difference is that while the eccentricities quickly rise to a small equilibrium value, the inclination in each system remains around the initial, near-coplanar values until the first close encounter occurs.  Inclinations are measured from the initial ecliptic plane. After the first encounter, the RMS inclination grows quickly through the first few encounters to a value of a few degrees.  Then the inclinations across the suite follow an evolutionary path similar to that of the eccentricities---with the typical RMS inclination of the system scaling as $i \propto  t^{1/3}$.  The evolution of inclination in each system can be seen in Fig. \ref{fig:loginc}.  Both RMS inclination and RMS eccentricity are also shown on a linear-log plot in Fig. \ref{fig:lininc} compared with the distributions of encounter and collision times.

To test whether or not this evolution changes with different initial conditions, we ran six suites of simulations with varying initial inclinations.  The inclination of each planet is chosen from a Rayleigh distribution where the mean value of the distribution increases by multiples of ten from $5\times10^{-5}$ to $5.0$ degrees.  We found consistent results.  Inclinations remain near the initial conditions up to the first encounter, followed by a steep rise over a factor of ten in time to an RMS inclination of around one degree. After that, the inclinations grow more slowly at the rates reported in the previous paragraph.

\begin{figure}
\includegraphics[width=\columnwidth]{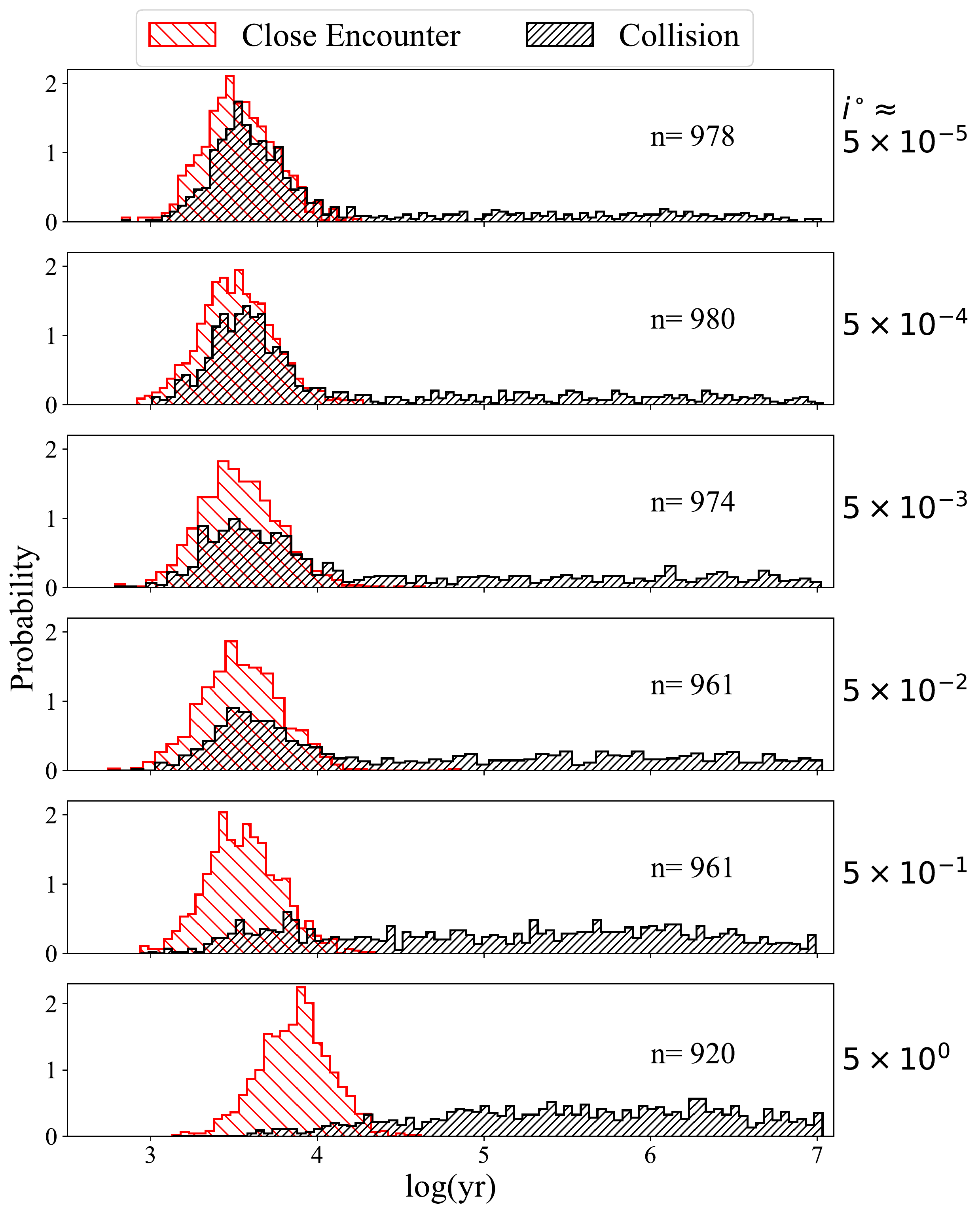}
\caption{Normalized histograms of encounter times and collision times for suites of simulations with increasing initial inclination (noted on the right in degrees).  The number of systems that had a planetary collision within the integration time of $10^{7}$ years is also noted.\label{fig:encountertocollision}}
\end{figure}

While the growth of inclination was similar across our range of starting values, different initial mutual inclinations did change the average lifetime of the systems.  Fig. \ref{fig:encountertocollision} shows that giving the suite a larger initial inclination distribution causes the population that experiences a prompt collision to diminish and the typical time to the first collision to grow.  Once the initial inclinations are of order one degree, we see that systems no longer have prompt collisions and the distribution of collision times becomes almost entirely detached from the distribution of encounter times.  The increase in collision time with increased inclination has also been reported in \citet{dawson2} and \citet{matsumoto3}.

\begin{figure}
\includegraphics[width=\columnwidth]{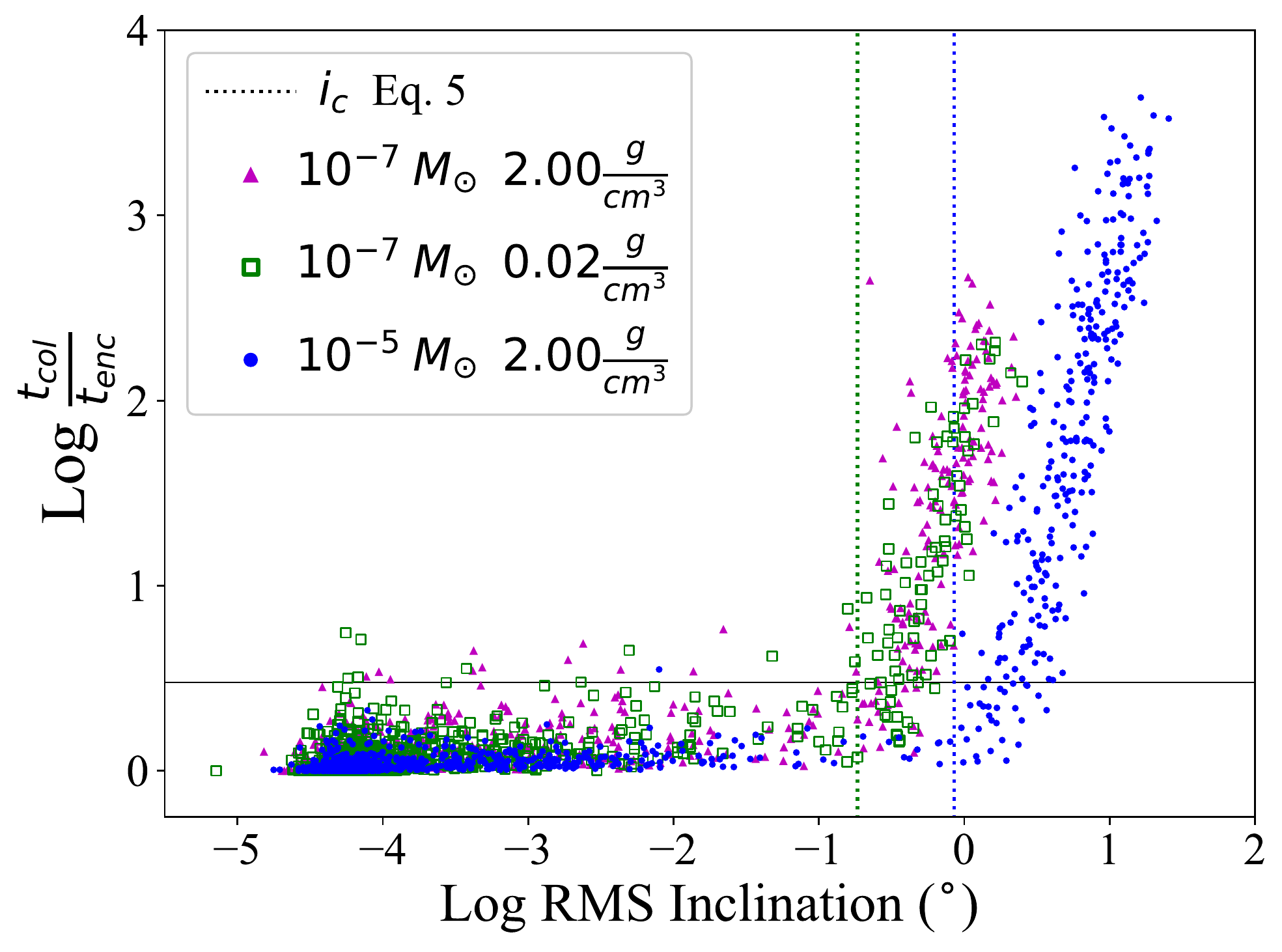}
\caption{The ratio of the collision time to the encounter time in a system is shown against the RMS inclination in log-degrees of the four planets.  The properties of each system is measured within the 100 years prior to the collision event (the time resolution of our analysis).  The three suites shown are the $\Delta=5$, near-circular, and near-coplanar systems with four planets of the given mass and density.  The solid line marks our threshold for long lived systems.  A bend occurs at the inclination that a majority of long lived systems reach before a collision.  In the $10^{-5} M_{\odot}$ suite, long lived systems have inclinations larger than 0.86 ($\sim10^{-0.07}$) degrees.  In the $10^{-7} M_{\odot}$ suites, long lived systems have inclinations greater than 0.18 ($\sim10^{-0.7}$) degrees.  The $10^{-7} M_{\odot}$ suites are integrated for $10^{6}$ years while the $10^{-5} M_{\odot}$ suite is integrated for $10^{7}$ years.  \label{fig:encountertime}}
\end{figure}

A final representation of the relationship between inclination and collision time is given in Figure \ref{fig:encountertime}.  For the initially near-coplanar suite, it shows the ratio of the collision to encounter time as a function of the system's RMS inclination near the time of the first planetary collision (our time resolution is 100 years).  A majority of systems with prompt collisions have inclinations $< 1^\circ$ near the time of the collision.  A striking feature appears at $\sim 1^\circ$ where systems begin to be long lived, $t_{col}/t_{enc}> 3$.  The long lived systems can have inclinations as large as 10$^\circ$.  The RMS inclination where the bend toward long lived systems occurs corresponds to the average ratio of the Hill Radius to the semi-major axis (the ``normalized'' Hill radius).  For our systems of $10^{-5} M_{\odot}$ planets this ratio is

\begin{equation}
i_{c}\simeq\frac{R_{H}}{a}\simeq\Bigl(\frac{m_{p}}{3M}\Bigr)^{\frac{1}{3}}\simeq0.86^{\circ}. \label{eq:inc}
\end{equation}

We see in Equation (\ref{eq:inc}) that the critical inclination for long lived systems (the normalized Hill radius) depends only on the mass of the planets and the central star---not on their densities or physical sizes.  In order to show the mass dependency, we run two suites of systems with equal planet masses of $10^{-7} M_{\odot}$.  The spacing, which depends on the planet mass, was kept at $\Delta=5$.  In the first suite, we keep the original density of 2.00  g/cm$^{3}$.  In the second, the planet radius is kept constant by changing the density to 0.02 g/cm$^{3}$.  The critical inclination from Eq. (\ref{eq:inc}) for $10^{-7} M_{\odot}$ planets is approximately 0.18$^{\circ}$.  Seen in Fig. \ref{fig:encountertime}, the inclination where long lived systems become prominent for both lower-mass suites corresponds to this predicted critical inclination---despite the factor of $10^{2/3}$ ($\simeq 4.6$) difference in planet radius.%  Future studies will explore this inclination criteria.

Also of note is the amount of long lived systems.  Using our criteria of the collision time being three times greater than the encounter time, 28\% of the $10^{-5} M_{\odot}$ systems are long lived.  Systems with the same density and $10^{-7} M_{\odot}$ planets are 35\% long lived.  (Systems that do not have a collision within the integration time are also considered long lived.)  However, in the low-mass/low-density suite, all the systems experience a collision within the integration time and only 8\% of systems were long lived.  In these systems the inflated planet size makes the planet fill a larger portion of its Hill sphere---giving them a larger collision cross section for each encounter.%  Thus fewer systems are long lived as discussed in Section \ref{sec:collision}.

These results show the interplay between the inclination of a system and the collision time.  Even in low-inclination systems, the inclination of a planet can grow through close encounters.  When the RMS inclination is larger than the ratio of the Hill radius to the orbital distance, the time of collision is no longer described by the size of the planet within the Hill sphere.  We see in our simulations that systems that initially have (Fig. \ref{fig:encountertocollision}) or that evolve to have (Fig. \ref{fig:encountertime}) the critical inclination have a distribution of collision times that is decoupled from the distribution of encounter times.  The three-dimensional nature of their orbits are realized once the mutual inclination is larger than the normalized Hill radius (Equation \ref{eq:inc}) and they are no longer strictly crossing.  The inclination can continue to increase through close encounters---further lengthening the system's lifetime.

\section{Collision Branching Ratios}\label{sec:branches}

\begin{figure}
\includegraphics[width=\columnwidth]{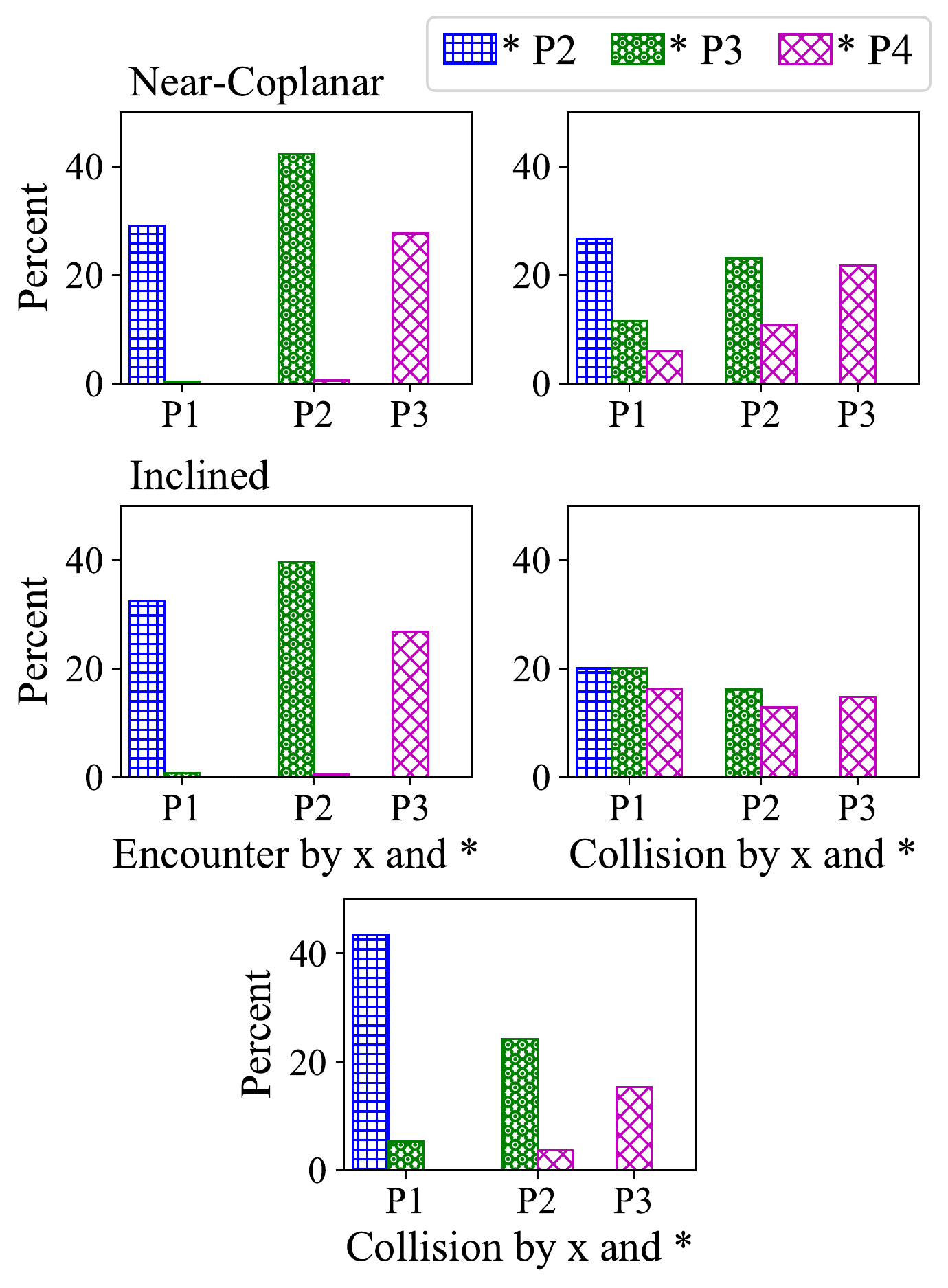}
\caption{The frequencies of which planets are involved in an instability event in the 1,000 systems with initial inclinations around $5\times10^{-5}$ degrees, \textit{top}, and around 5 degrees, \textit{middle}.  Planets are labeled sequentially by their initial position from innermost to outermost.  \textit{Left} shows which planet on the x-axis was encountered by the planet of the corresponding color during each system's first close encounter.  \textit{Right} similarly shows which planets were involved in each system's first collision.  \textit{Bottom} shows the frequencies of planet pairs that are involved in the collision in the inclined systems when the planets are renamed by their position from inner to outer within 100 years before the collision. \label{fig:branch}}
\end{figure}

The difference between systems that undergo a prompt collision and long lived systems can also be seen in the planets that are involved in the collision.  In all of our suites, the frequency with which planet pairs are involved in the first close encounter is similar.  Over 99.5\% of systems have the first encounter between neighboring planets.  The most common encounters (over 40\%) are between the middle two planets.  Encounters between the inner two and outer two planets are equally likely, with each occurring in roughly 30\% of the systems.  These frequencies can be seen in the left of Fig. \ref{fig:branch} for two suites of simulations with different initial inclinations---$5\times10^{-5}$ degrees and $5.0$ degrees.

The right side of Fig. \ref{fig:branch} shows the frequencies of planet pairs that are involved in the first collision.  We show that the frequencies are affected by how long the systems typically survive following the first encounter.  In our near-coplanar suite, the first collision occurs between nearest neighbors about 75\% of the time---a similar rate to the number of systems with prompt collisions.  Prompt collisions do not always occur between the same planets that had the first close encounter, but it remains more likely for nearest neighbors to collide.

The initially inclined systems, on the other hand, are almost entirely long lived as shown in Fig. \ref{fig:encountertocollision}.  The percent of systems with a collision between nearest neighbors is around 50\%---significantly less than the frequency in the near-coplanar suite.  In Fig. \ref{fig:branch} we show that the probability of any two planets colliding is approximately one-sixth ($\approx 16\%$) with only a slightly higher probability for the innermost planets.  The first collision in these systems occurs with roughly equal probability between any planet pair, regardless of their initial position.  Lastly, in Fig. \ref{fig:branch} we recover collision frequencies similar to that of the non-coplanar suite when the planets are renamed in each system from inner to outer within 100 years prior to the collision event.  The collision occurs between planets which are neighbors within 100 years before the collision in approximately 90\% of the systems. We expect that the other 10\% of systems where non-neighbors collide have high eccentricities.

These results suggests that the orbits of the planets in long lived systems are significantly mixed from their initial order before a collision event. When we examine individual systems we see this is the case.  Systems remain in chaotic evolution between the first encounter and collision.  During this period, the planets experience multiple changes in semi-major axis which are often larger than a 10\% change. 

\section{Conclusion}\label{sec:conclusion}

We studied the distributions of instability times in systems of four, equally-spaced, Neptune-like planets.  We investigated the difference between measuring instability time as the time from initial conditions to the first close encounter and as the time to the first planet-planet collision.  Our findings, and their implications, are summarized as follows:

\begin{enumerate}
\item The distribution of encounter times for systems of a given orbital separation is approximately log-normal and it spans an order of magnitude in the number of orbits of the innermost planet.  The encounter time distribution is independent of the innermost planet's semi-major axis as expected from dynamical scaling relationships (the Hill sphere being proportional to the orbital distance). 
\item After experiencing a close encounter, an unstable system with non-zero mutual inclination can persist without a collision for a much longer period of time. The ratio of the collision time to the encounter time can be a few orders of magnitude, $t_{col}/t_{enc}\gg 1$.
\item In a long lived system, which is dynamically unstable but has not had a collision event, the first few close encounters set the RMS eccentricity and inclination of the system onto a new evolutionary path where the RMS eccentricity grows as $e \propto t^{1/6}$ and the RMS inclination grows as $i \propto  t^{1/3}$.
\item If planets in a non-coplanar system fill a majority of their respected Hill spheres the collision time is similar to the encounter time and follows approximately the same distribution as encounter time.  However, when the planets are much smaller than their Hill Sphere the probability of a collision is decreased, and the system has more time to excite inclinations through multiple close encounters.
\item Systems that either initially have, or evolve to have, mutual inclinations that are larger than the average ratio of each planet's Hill Radius to its semi-major axis (Eq. \ref{eq:inc}), do not experience prompt collisions.  The average time of a collision in a system with raised inclination is much longer.  Systems with significant inclinations have $t_{col}/t_{enc}\ge 3$ (our chosen cutoff for a prompt collision) with some not experiencing a collision for $t_{col}/t_{enc}\ge 1000$.
\item  In systems with prompt collisions, planets that are initially nearest neighbors are most likely to be involved in the collision.  Long lived systems experience ongoing changes to the orbits of the planets and exhibit no preference as to which planets collide.  However, when the reordering of planetary orbits during the system's dynamical evolution is accounted for, nearest neighbor collisions are again preferred.
\end{enumerate}

These results have some implications for the stability of systems similar to TRAPPIST-1.  From the parameters in \citet{grimm}, we find the spacings of the TRAPPIST-1 system range from $\Delta\approx 6.8 - 13.4$.  Using the relationship for Earth-mass systems in \citet{obertas}, the smallest spacing yields an expected close encounter time of $10^{5.3}$ orbits of the innermost planet.  For TRAPPIST-1 this is only about 800 years---much less than the 7.6 Gyr age of the system \citep{burgasser}.  The observed orbital resonances \citep{luger, matsumoto} must be invoked to explain TRAPPIST-1's long term stability (something we do not consider in this work).

However, consider a TRAPPIST-1-like system that has its inner planet at 1.0 AU and is not protected by resonances. This system could have an encounter timescale on the order of a few Myr (if it were in the rightmost tail of the encounter distribution---see Fig. \ref{fig:gauss}).  Our results suggest that such a system could survive multiple orders of magnitude longer following a close encounter.  If that system had inclinations above $\simeq1.2^{\circ}$, from Eq. (\ref{eq:inc}) with TRAPPIST-1 star and planetary masses, it could survive without a collision for Gyr timescales.  The range of typical mutual inclinations for \textit{Kepler} multis encompasses this critical inclination---$1.0^{\circ}<i_{\textit{Kepler}}<2.2^{\circ}$ \citep{tremaine, fang2, fabrycky}.  We show in Fig. \ref{fig:lininc} that long lived systems will be observed to have large eccentricities and inclinations.  Furthermore, systems with slightly larger separations, with encounter timescales of 10 to 100 Myr, could survive for the lifetime of a typical G-type star.  Thus, the systems that we observe today, and that we initially assume are stable given the age of the host star, may in fact have long ago experienced the encounter that would traditionally mark them as unstable.  It remains unclear the full ramifications of this finding.  However, when interpreting observational data, it does suggest that some caution be exercised when constraining the orbital parameters of a system by invoking dynamical stability.

\section*{Acknowledgements}

JHS and DRR acknowledge support from the College of Sciences at the University of Nevada, Las Vegas, the Center For Interdisciplinary Exploration and Research in Astrophysics (CIERA) at Northwestern University, and NASA grants NNX16AK32G and NNX16AK08G.  All simulations were supported by the Quest high performance computing facility at Northwestern University.  We acknowledge that the study resulting in this publication was assisted by grants from the WCAS Undergraduate Research Grant Program which is administered by Northwestern University's Weinberg College of Arts and Sciences.

%%%%%%%%%%%%%%%%%%%%%%%%%%%%%%%%%%%%%%%%%%%%%%%%%%

%%%%%%%%%%%%%%%%%%%% REFERENCES %%%%%%%%%%%%%%%%%%

% The best way to enter references is to use BibTeX:

\bibliographystyle{mnras}
\bibliography{references} % if your bibtex file is called example.bib

% Don't change these lines
\bsp	% typesetting comment
\label{lastpage}
\end{document}